# Unraveling the PFAS helix: A statistical approach

*Pranoy Ray[1,2], Haden Cavalli[3], Gashaw Bizana[4], Andrew R. Castillo[4], Shubham Vyas[3], Ronald L. Siefert[5], Surya R. Kalidindi[1,2], Manoj Kolel-Veetil[6]**

[1] George W. Woodruff School of Mechanical Engineering, Georgia Institute of Technology, Atlanta, USA
[2] School of Computational Science and Engineering, Georgia Institute of Technology, Atlanta, USA
[3] Department of Chemistry, Colorado School of Mines, Golden, USA
[4] Multiscale Technologies Inc., Seattle, USA
[5] United States Naval Academy, Annapolis, USA
[6] United States Naval Research Laboratory, Washington DC, USA

*Corresponding author email: manoj.k.kolel-veetil.civ@us.navy.mil

## ABSTRACT

The extreme persistence of per- and polyfluoroalkyl substances (PFAS) in the environment is rooted in their three-dimensional molecular conformation. The helical twist adopted by perfluoroalkyl chains due to hyperconjugation involving their recalcitrant C-F bonds governs their resistance to degradation, yet a quantitative, continuous metric for helicity has remained absent. Here we introduce a data-driven framework that quantifies backbone helicity using three statistical descriptors: void-state (binary occupancy) spatial autocorrelations, local backbone principal component analysis, and persistent homology. We further validate each statistical descriptor against geometric dihedral benchmarks and DFT-calculated Vibrational Circular Dichroism (VCD) spectra. The framework is initially established on a homologous series of perfluorocarboxylic acids ($^{F}C_2$ through $^{F}C_{16}$) and their hydrogenated analogues, then extended across perfluorosulfonic acids, fluorotelomer alcohols, polyfluoroalkyl hexanoic acid analogues, and longer-chain PFOA and PFOS analogues. Agreement between statistical, geometric, and spectroscopic definitions of helicity is established for PFCAs and extended to structurally distinct subfamilies, including PFOA analogues of varying fluorine content and perfluorosulfonic acids, demonstrating that the descriptors are robust to changes in headgroup chemistry and fluorine substitution pattern. The framework provides a chemistry-agnostic toolset for incorporating backbone conformation into predictive models of PFAS environmental fate and degradation reactivity.

## 1. INTRODUCTION

Per- and polyfluoroalkyl substances (PFAS)[1-5] are a class of over 21,000 synthetic organofluorine compounds[6] whose widespread use has led to their emergence as ubiquitous and persistent global pollutants. Termed "forever chemicals", their defining characteristic is an extraordinary resistance[7,8] to chemical, biological, and thermal degradation. This recalcitrance is primarily attributed to the exceptional thermodynamic stability of the carbon-fluorine (C-F) bond and the dense sheath of electronegative fluorine atoms that sterically shield the carbon backbone from nucleophilic attack. Consequently, PFAS bioaccumulate in ecosystems and living organisms, where they are linked to a range of adverse health outcomes[9], creating an urgent global challenge for environmental[10] remediation.

Prevailing degradation strategies[1,11–14] have logically focused on supplying sufficient energy via advanced oxidation and reduction, electrochemical methods, or plasma treatment[1] to overcome the high activation barrier for C-F bond cleavage. However, this focus on bond energetics often overlooks a more subtle, yet equally critical, determinant of reactivity: the three-dimensional conformation of PFAS. A fundamental structural dichotomy exists between perfluoroalkanes and their hydrogenated hydrocarbon counterparts. While n-alkanes adopt a planar, all-anti (zigzag) geometry, the steric and electrostatic repulsion between the larger, highly electronegative, fluorine atoms induce helical twisting in perfluoroalkyl chains[15], with nascent helical[13-15] character appearing in chains as short as three carbons and pronounced compact helical twists emerging in chains longer than five carbons. This helical conformation is stabilized by favorable intramolecular dipole interactions and hyperconjugation[16].

Recent computational studies[17,19] have proposed that this helical structure is mechanistically coupled to PFAS degradation through a linearization-facilitated pathway. Reductive degradation pathways, which are among the most promising strategies for defluorination, are often initiated by the capture of an excess electron. This event has been proposed to form a transient radical anion that is electronically unstable in the compact helical conformation, such that electron capture may induce a conformational transition[17] from a helical toward a more linearized state. This linearization has been hypothesized to facilitate degradation by exposing the carbon backbone and altering C-F bond strengths, thereby lowering the barrier to bond scission. This proposed "linearization-facilitated degradation" mechanism is supported by converging computational and experimental evidence across several contexts[17-22], including surface-mediated reactions where substrates such as electrochemical anodes or zerovalent metals act as electron donors; however, direct experimental confirmation of the transient radical-anion intermediate remains an open challenge.

Despite its clear importance in governing PFAS stability and reactivity, helicity remains a largely qualitative concept, typically described by isolated dihedral angles that fail to capture the global topology of the molecular fold. The absence of a robust, continuous, and quantitative metric for helicity is a major gap in the field. Such a descriptor is essential for building the predictive Quantitative Structure-Property Relationship (QSPR) models[23-25] needed to elucidate the foundational mechanisms of PFAS persistence and thereby guide the rational design of novel chemical and biological degradation technologies. To address this gap, we hypothesized that a quantitative measure of helicity could be extracted from atomic backbone coordinates using chemistry-agnostic feature engineering protocols. To rigorously test this, we directly compare a homologous series of perfluorinated carboxylic acids (PFCAs, ${}^{F}C_2$-${}^{F}C_{16}$) with their corresponding hydrogenated n-alkyl carboxylic acid counterparts, allowing us to isolate the structural consequences of fluorination.

We employ three complementary statistical approaches[25-28], all operating exclusively on the sorted carbon backbone coordinates (Figure 1). The void-state voxelization and two-point spatial autocorrelation pipeline[32] establishes a global shape descriptor which cleanly separates fluorinated from hydrogenated chains across all PFAS subfamilies, confirming the dichotomy in backbone morphology. Local principal component analysis of each molecule's backbone coordinates extracts the explained variance ratio as a per-molecule out-of-plane twist measure, providing a within-class helicity gradient. Persistent homology of the arc-length-normalized backbone generates topological fingerprints whose principal component scores prove to be the strongest continuous helicity descriptor, resolving chain-length-dependent helical progression even within the fluorinated series. The physical validity of all three descriptors is established by cross-validation against C-C-C-C backbone dihedral deviation from planarity and median F-C-C-F dihedral distributions, while DFT-calculated VCD[33] spectra provide theoretical validations with quantum chemistry.

# 2. METHODOLOGY

## 2.1 DFT CALCULATIONS

A homologous series of perfluorinated carboxylic acids from perfluoroethanoic acid (${}^{F}C_{2}$) to perfluorohexadecanoic acid (${}^{F}C_{16}$) was selected to systematically investigate chain-length-dependent helicity. To create a rigorous control group isolating the structural effects of fluorination, an analogous series of hydrogenated n-alkylcarboxylic acids (${}^{H}C_{2}$ through ${}^{H}C_{16}$) was modeled using identical computational parameters. This chemical series provides an ideal testbed because: (i) helical propensity increases monotonically with chain length, (ii) structural differences are isolated to chain length while maintaining constant head group chemistry, and (iii) prior NMR[34] and crystallographic studies[35,36] confirm that this structural system is well-characterized, providing a known-geometry foundation for the present computational framework. Next, Density Functional Theory (DFT)[33-37] geometry optimizations and frequency calculations were conducted using the ORCA 6.0[42] software package. The M06-2X density functional with the 6-31+G(2d,p)[43,44] basis sets were used for this study. The choice of methods, particularly the M06-2X functional and split-valence double-zeta basis set, is supported by Giroday *et al.*[45] and Hidalgo *et al.*[46] which demonstrated good PFOA benchmarking values compared to other common PFAS DFT methods. All optimizations were performed with tight SCF convergence criteria and confirmed as true minima via subsequent frequency calculations. Full DFT computational details are provided in Supplementary Information S1. All descriptors in this work are evaluated on the single DFT free-energy-minimum (helical) conformer of each molecule. This single-conformer choice is deliberate: a well-defined, reproducible geometry per molecule is required to establish an unambiguous ground-truth ordering of helicity against chain length and fluorination. Prior conformational-distribution studies of perfluoroalkyl substances show that the helical family dominates the low-energy landscape[19], so the minimum-energy conformer is representative of the Boltzmann-weighted ensemble over the chain lengths considered. Each descriptor is nonetheless ensemble-ready and could be Boltzmann-averaged over a sampled conformer set; conformer-averaged descriptors are identified as a natural extension in Section 4.

## 2.2 FEATURE ENGINEERING

All three statistical descriptors are derived exclusively from the ordered carbon backbone coordinates of each molecule, extracted via graph-theoretic identification of the longest carbon path in the molecular connectivity network and oriented from the carboxylate head group outward (hereafter referred to as the sorted backbone), ensuring that all descriptors operate on identical geometric representations. This ensures that all descriptors operate on identical geometric representations and encode no explicit chemical identity information beyond carbon connectivity. The complete pipeline from backbone extraction to descriptor output is illustrated in Figure 1.

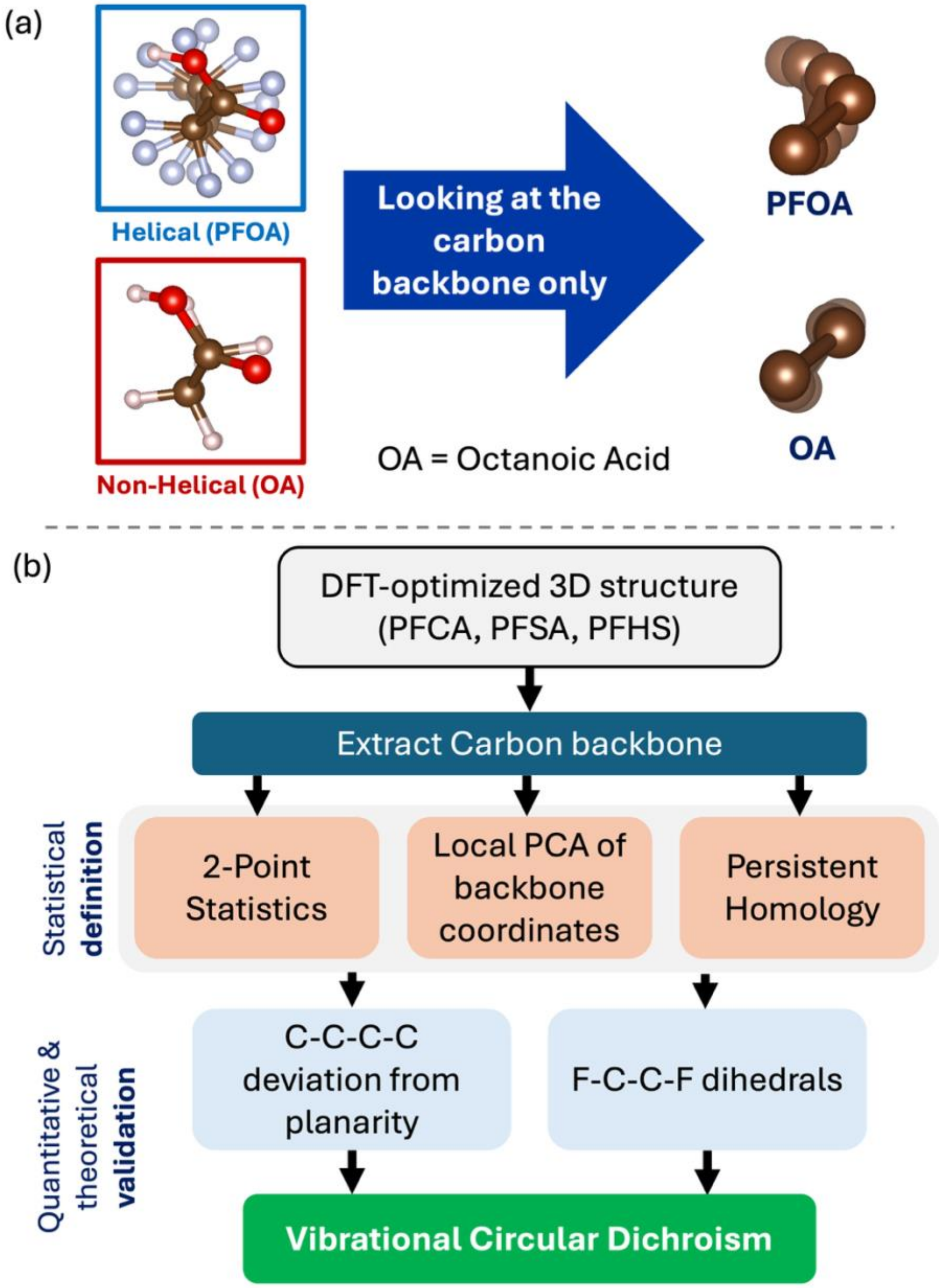

**Figure 1:** Feature engineering workflow for quantification of PFAS backbone helicity. (a) The carbon backbone is isolated from the full DFT-optimized structure, illustrated here for PFOA (helical) and octanoic acid (planar). (b) Three statistical descriptors are derived exclusively from the sorted carbon backbone coordinates. The void-state voxelization and two-point spatial autocorrelation pipeline (left branch) generate a global backbone shape descriptor; Spatial PC2 separates fluorinated from hydrogenated classes. Local principal component analysis (center branch) yields Local PC3 EVR as a per-molecule out-of-plane twist measure. Persistent homology of the arc-length-normalized backbone (right branch) produces persistence-image fingerprints whose first principal component (PH PC1) acts as the strongest continuous helicity descriptor. All three descriptors are cross-validated against C-C-C-C dihedral benchmarks and DFT-calculated VCD spectra.

### 2.2.1. Voxelization & Spatial Autocorrelations

A global backbone shape descriptor is derived by applying the molecular voxelization and two-point spatial autocorrelation pipeline[28,47] established in Ray et al.[32] to the carbon backbone coordinates pooled across all PFAS subfamilies. Two adaptations are made for the present application. First, voxelization is performed exclusively on the sorted carbon backbone atom coordinates rather than the full molecular structure; this restriction simplifies the input and ensures the descriptor remains chemistry-agnostic, avoiding bias introduced by the highly polar C-F bonding geometry. Second, the scalar field defined on the voxel grid is a binary void-state occupancy field: $\rho(s) = 1$ if voxel $s$ intersects the van der Waals volume of any backbone carbon atom and $\rho(s) = 0$ otherwise. This representation encodes the spatial occupancy pattern of the carbon backbone with no explicit chemical identity information beyond carbon connectivity itself.

From this field, two-point spatial autocorrelations $f_r^0$ are computed efficiently via Fast Fourier Transforms with non-periodic boundary conditions, following Ray et al.[32]. The autocorrelation features are standardized to zero mean and unit variance and reduced via global PCA performed once across all subfamilies. Because PCA is fitted globally, all molecules are projected into a common coordinate system, enabling direct cross-class structural comparison. For screening, this basis is frozen: the standardization parameters and the principal-component loading vectors are fit once on the training set and stored. A new molecule is standardized with the stored parameters and projected onto the frozen loadings (an out-of-sample transform, not a refit), so its Spatial PC scores lie on the same axes as the training set and are directly comparable. Within this global PCA decomposition, the first principal component is dominated by chain length and overall molecular extent, as expected for a global autocorrelation of the backbone occupancy field. The second principal component (Spatial PC2) captures the cross-sectional morphological difference between helical and planar backbone geometries and cleanly separates fluorinated from hydrogenated chains across all subfamilies. Spatial PC2 is therefore the void-state descriptor used in all subsequent cross-validation analyses. The explained-variance ratios of the leading components are Spatial PC1 = 20.6%, Spatial PC2 = 5.9%, and Spatial PC3 = 5.2% (cumulative 31.7% for the first three). The eigenvalue spectrum shows a pronounced elbow after PC1, consistent with the chain-length dominance of Spatial PC1, and a secondary elbow after PC2, supporting the use of Spatial PC2 as the class-separating, helicity-relevant axis; the full scree plot is provided in Supplementary Figure S7.

### 2.2.2. Local Backbone Principal Component Analysis

As a second backbone shape descriptor, PCA is applied locally to the three-dimensional carbon backbone coordinates of each molecule individually. For a molecule with N backbone carbons, PCA decomposes the coordinate variance into three orthogonal components: PC1 captures the primary chain axis (elongation), PC2 captures in-plane bending, and PC3 captures out-of-plane twist. The PC3 explained variance ratio (Local PC3 EVR) serves as a per-molecule scalar measure of out-of-plane backbone variance: a strictly linear chain yields Local PC3 EVR near zero, while a chain with out-of-plane backbone distortion distributes a measurable fraction of the total coordinate variance into the third component. Local PC3 EVR provides a direct, per-molecule geometric proxy for helical character without requiring any global reference frame.

### 2.2.3. Persistent Homology Fingerprints

A pose-normalized persistent-homology fingerprint was computed for each PFAS molecule from the carbon-backbone geometry. Persistent homology (PH) quantifies the lifespan of k-dimensional topological features, such as connected components, loops, and cavities, within a growing simplicial complex and provides a multiscale summary of molecular geometry[44-48]. As with the void-state pipeline, the input point cloud consisted exclusively of the ordered carbon backbone coordinates, ensuring that all statistical descriptors operated on identical geometric representations. The full procedure is detailed in Supplementary Information S2. Briefly, carbon atoms were extracted from each structure and ordered along the backbone using a minimum-spanning-tree-based traversal. The resulting backbone curve was recentered and rescaled to unit mean inter-carbon distance. It was then treated as a piecewise-linear curve, resampled at 500 evenly spaced positions along its cumulative path length, and aligned to a canonical PCA/SVD coordinate frame. The canonicalized backbone was transformed to a circular angular embedding, $(\cos\theta,\ \sin\theta,\ z)$, where $\theta$ is the cylindrical angle about the selected backbone axis. Persistent homology was computed in HomCloud[51] using an alpha filtration, and the $H_1$ persistence diagram was converted to a persistence image weighted by lifetime and convolved with a Gaussian kernel, yielding a fixed-length vector representation[53]. PCA was applied to the persistence-image vectors across all molecules, and the resulting PH principal-component scores were used as molecular descriptors; PH PC1 was retained as a per-molecule scalar descriptor for downstream analysis. As with the spatial descriptor, the persistence-image standardization and PCA loadings are frozen after fitting on the training set; a new molecule's persistence-image vector is projected onto this fixed basis to obtain its PH PC1 score, without refitting the PCA. This preprocessing reduces sensitivity to rigid-body translation and rotation, standardizes the number of sampled backbone points across molecules, and normalizes local inter-carbon length scale, while preserving information related to backbone shape and angular organization.

### 2.2.4. Dihedral Geometric Benchmarks

Two dihedral-based geometric benchmarks are computed for each molecule and used exclusively for cross-validation of the statistical descriptors; they are not used as input features. First, all C-C-C-C backbone dihedral angles along the sorted carbon chain are extracted using RDKit[54]. For a chain of $N$ backbone carbons, this yields $N - 3$ dihedral angles. The deviation from planarity is defined as $\min(\theta, 180 - \theta)$ applied to the mean backbone dihedral angle theta, where a value of zero corresponds to the all-anti planar conformation and increasing values indicate progressive departure toward gauche arrangements consistent with helical twisting. Second, for fluorinated molecules, all F-C-C-F dihedral angles across every consecutive C-C bond in the backbone are extracted; for hydrogenated analogues, the equivalent H-C-C-H dihedrals are used. The median of these distributions is computed per molecule. In a helical perfluoroalkyl chain, the gauche preference of adjacent fluorines drives F-C-C-F dihedrals toward approximately 60°-80° (Figure 3b), in contrast to the near-180° anti values observed in linear hydrogenated analogues (Figure 3b).

## 3. RESULTS & DISCUSSION

### 3.1. Separation of Fluorinated and Hydrogenated Classes by Backbone Shape

Figure 2a shows the Spatial PC1 vs. Spatial PC2 projection of the void-state spatial autocorrelation PCA for the PFCA homologous series. As described in Section 2.2.1, separation of fluorinated from hydrogenated chains occurs along Spatial PC2. Fluorinated molecules from all subfamilies (PFCA, PFOA, PFOS, PFSA, PFHS, FTOH) occupy a region of negative Spatial PC2[⊥] that is distinct from the hydrogenated PFCA analogues, which cluster at near-zero and positive Spatial PC2 values. Within the fluorinated cluster, the spread along Spatial PC1 tracks chain length without resolving a helicity gradient. A monotonic chain-length-dependent helicity gradient within the fluorinated subset is not resolved by Spatial PC2 alone; this gradient is recovered by the descriptors introduced in the following sections.

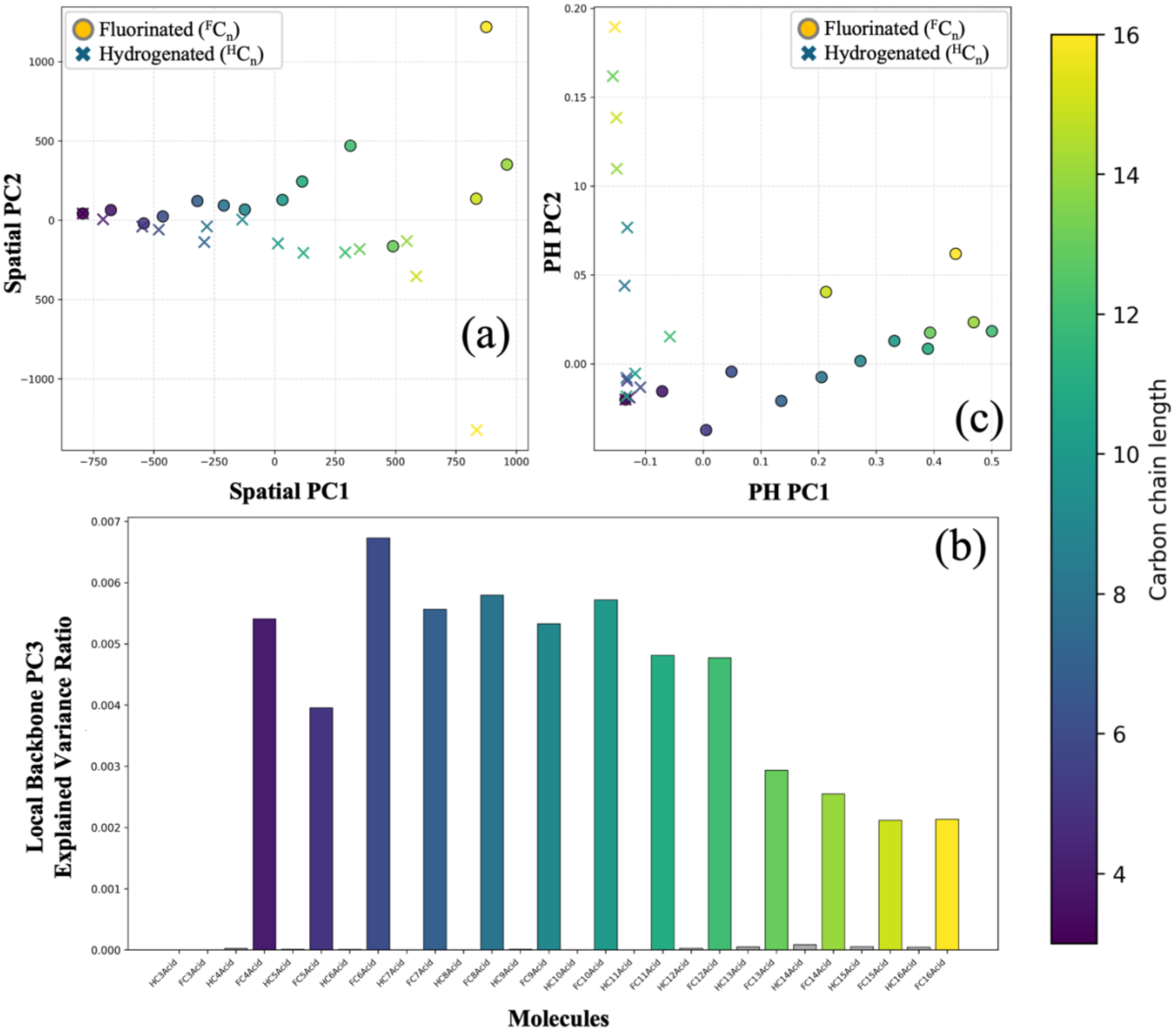

[⊥] the sign of PC2 is a PCA convention reflecting the direction of the loading vector; the key result is the separation between classes, not the sign itself

**Figure 2:** Statistical descriptors of carbon backbone shape across the PFCA homologous series. (a) Spatial PC1 vs. Spatial PC2 from void-state two-point spatial autocorrelations. The two molecular classes separate along PC2; PC1 is dominated by chain extent. Fluorinated molecules (filled circles) and hydrogenated analogues (crosses) are colored by carbon chain length. (b) Local backbone PC3 explained variance ratio (Local PC3 EVR), with molecules ordered $^{H}C_n$ then $^{F}C_n$ within each chain length. Grey bars: hydrogenated analogues; colored bars: fluorinated acids (colormap scales by chain length). Fluorinated chains show systematically larger Local PC3 EVR than their hydrogenated analogues at every chain length above $C_5$. Because Local PC3 EVR is an explained-variance ratio, it does not increase monotonically with chain length within the fluorinated series: it is largest for short-to-mid chains and declines for the longest chains as PC1 (chain elongation) captures a growing fraction of the total variance. The monotonic chain-length helicity gradient is instead recovered by PH PC1 (panel c). (c) Persistent-homology principal component space (PH PC1 vs. PH PC2). Fluorinated acids $^{F}C_2$ through $^{F}C_{16}$ form a monotonically ordered sequence along PH PC1 with increasing chain length; hydrogenated analogues cluster at low PH PC1 regardless of chain length.

Figure 2b shows Local PC3 EVR for the PFCA series, with hydrogenated ($^{H}C_n$) and fluorinated ($^{F}C_n$) molecules paired by chain length. Within each chain length, the fluorinated acid shows systematically larger Local PC3 EVR than its hydrogenated analogue. Because Local PC3 EVR is a variance ratio, it does not rise monotonically with chain length within the fluorinated series; it peaks for short-to-mid chains and decreases for the longest chains, since PC1 (elongation) absorbs an increasing share of the coordinate variance as the chain extends even as the absolute out-of-plane distortion grows. Local PC3 EVR therefore functions as a robust per-molecule discriminator between the helical fluorinated and planar hydrogenated classes rather than as a within-class chain-length gradient, the latter being provided by PH PC1. Hydrogenated analogues maintain near-zero values throughout, confirming their planar all-anti conformations.

In the persistent-homology space (Figure 2c), hydrogenated analogues cluster near zero PH PC1 regardless of chain length, while fluorinated acids form a monotonically ordered trajectory along PH PC1 with increasing chain length. The arc-length normalization in the PH pipeline removes the chain-length contribution that dominates Spatial PC1, leaving the residual loop and cavity topology of the helical fold as the dominant source of variance. Short fluorinated chains ($^{F}C_2$, $^{F}C_3$) cluster near their hydrogenated analogues, consistent with their predominantly extended conformations, while $^{F}C_8$ through $^{F}C_{16}$ form a well-resolved monotonic sequence. Across all three descriptors, helicity in PFAS is not a binary switch but a continuous gradient that increases progressively with chain elongation.

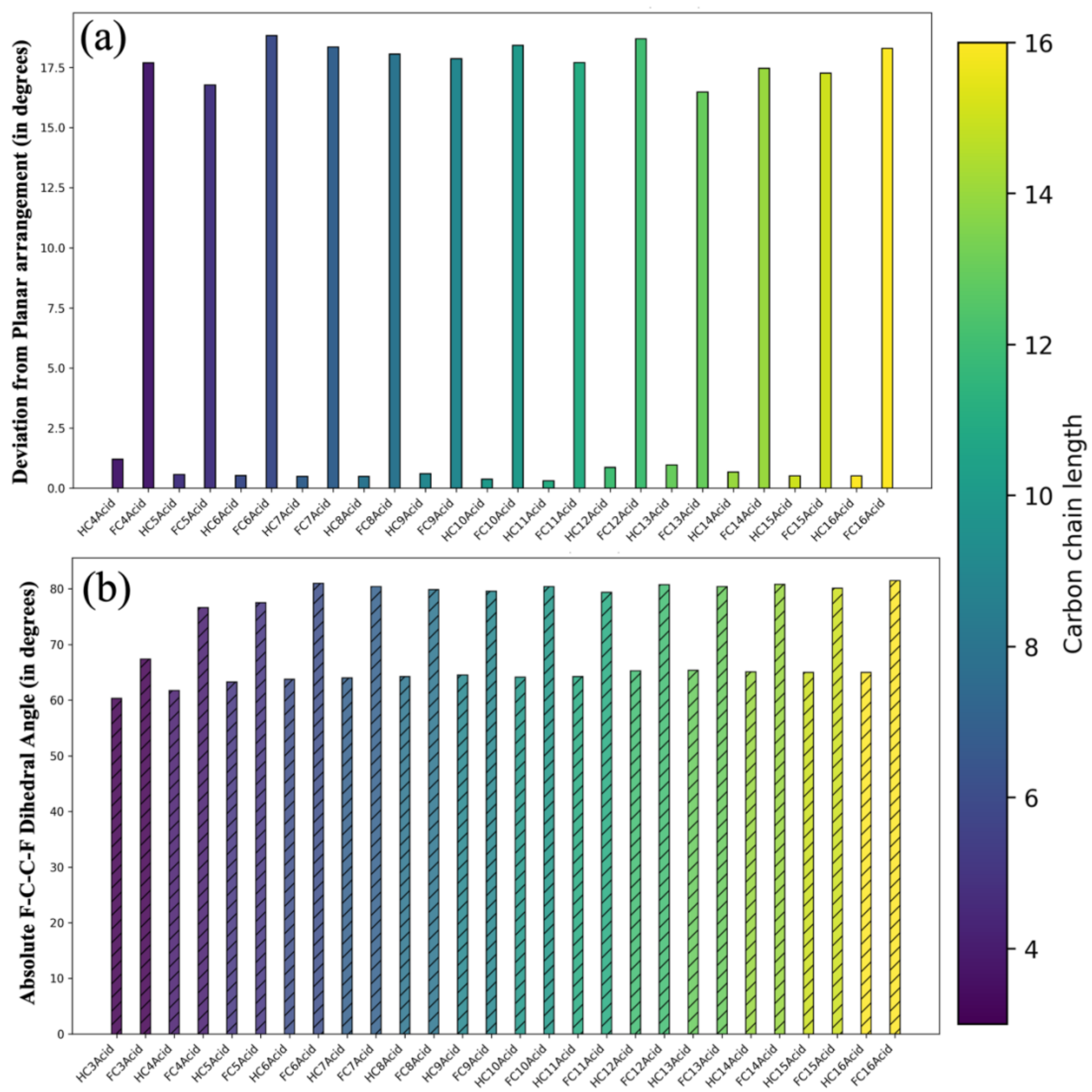

**Figure 3:** Geometric benchmarks for backbone helicity across the PFCA homologous series. (a) C-C-C-C backbone dihedral deviation from planarity. Fluorinated acids (colored by chain length) show near-zero deviation for $FC_2$ and $FC_3$, consistent with predominantly extended conformations in the shortest chains, and rise monotonically from $FC_4$ onward, reaching 17°-19° for $FC_{10}$ through $FC_{16}$. Hydrogenated analogues (grey) remain below 1.5° regardless of chain length. (b) Median absolute F-C-C-F dihedral (fluorinated) and median H-C-C-H dihedral (hydrogenated) for the PFCA series. Fluorinated chains cluster in the 60°-80° gauche range for all chain lengths above $C_4$, reflecting the gauche preference of adjacent fluorines that drives the helical backbone[13-15] geometry; hydrogenated chains maintain near-180° anti values throughout.

### 3.2. Geometric Validation of Statistical Descriptors

Having established the three statistical descriptors, we next evaluate whether they track physically meaningful geometric measures of backbone helicity. The two benchmarks, namely the C-C-C-C deviation from planarity and the median F-C-C-F dihedral, are computed as described in Section 2.2.4 and used here as reference measures only.

Figure 3 shows the two geometric benchmarks for the PFCA series. The deviation from planarity (Figure 3a) rises monotonically with chain length for the PFCA homologous series, reaching 17°-19° for $^{F}C_{10}$ through $^{F}C_{16}$, while all hydrogenated analogues remain below 1.5° regardless of chain length. The same monotonic rise with chain length, at comparable amplitudes, is reproduced in the PFOA, PFOS, PFSA, FTOH, and PFHS subfamilies, confirming the generality of the backbone twist across headgroup chemistries. The per-subfamily plots are provided in Supplementary Figures S12, S15, S18, S21, S24, and S25. The median F-C-C-F dihedral (Figure 3b) for PFCA chains lies in the 60°-80° range, consistent with the gauche conformational preference of adjacent fluorines that is the chemical origin of the helical twist. Hydrogenated analogues show median H-C-C-H values near 175°-180°, consistent with the all-anti conformation of linear alkyl chains established by conformational analysis[13-15]. The narrow spread of F-C-C-F values across subfamilies confirms that the gauche fluorine preference is a class-wide feature. The same gauche preference, with F-C-C-F medians in the 60°-80° range, is observed across all fluorinated subfamilies in the full multi-class dataset (Supplementary Section S4).

Cross-validation of the statistical descriptors against these geometric benchmarks is summarized in the Pearson $r$ correlation matrix (Figure 4). Individual scatter plots for each descriptor-benchmark pair are provided in Supplementary Figures S27 through S31. Within the PFCA series, PH PC1 shows the strongest Pearson correlations with both geometric benchmarks: a strong positive correlation with deviation from planarity ($r = +0.84$) and a positive correlation with the median F-C-C-F dihedral ($r = +0.69$, computed on fluorinated molecules only, whose gauche angle increases modestly from ≈60° to ≈80° with helical character). Local PC3 EVR shows moderate positive correlations with both benchmarks, consistent with its direct geometric interpretation as an out-of-plane variance fraction. Spatial PC2 shows weaker correlations, reflecting the fact that it resolves the class-level dichotomy but not the within-class chain-length gradient. Across the full multi-class dataset, the pattern is preserved: PH PC1 remains the strongest helicity proxy against geometric benchmarks, demonstrating that the descriptor generalizes across headgroup chemistries and chain architectures. Because the pooled PFCA dataset contains two well-separated classes (helical fluorinated and planar hydrogenated), part of the correlation in Figure 4 reflects between-class separation rather than a within-class gradient. Recomputing within the fluorinated cluster alone, the correlation of every descriptor with the C-C-C-C deviation-from-planarity benchmark drops sharply (Spatial PC2: +0.64 → +0.13; Local PC3 EVR: +0.92 → +0.57; PH PC1: +0.84 → +0.07). This drop reflects a limitation of that benchmark rather than of the descriptors: the C-C-C-C deviation is undefined for the shortest chains ($C_2$, $C_3$) and saturates at approximately 17-19° from $C_4$ onward (Figure 3a), leaving it with little dynamic range to resolve the within-series gradient. Benchmarks that retain dynamic range across the fluorinated series behave differently: PH PC1 remains the most strongly correlated descriptor with the median F-C-C-F dihedral within the fluorinated cluster ($r = +0.69$, versus +0.55 for Spatial PC2 and +0.52 for Local PC3 EVR), and the independent VCD observable (Figure 6) corroborates the same helical gradient. The continuous within-class resolution PH PC1 provides, beyond the point where the coarse dihedral benchmark saturates, is precisely its intended advantage. We therefore note explicitly that the pooled correlations in Figure 4 include a class-separation contribution; full pooled and within-cluster correlations are tabulated in Supplementary Table S3.

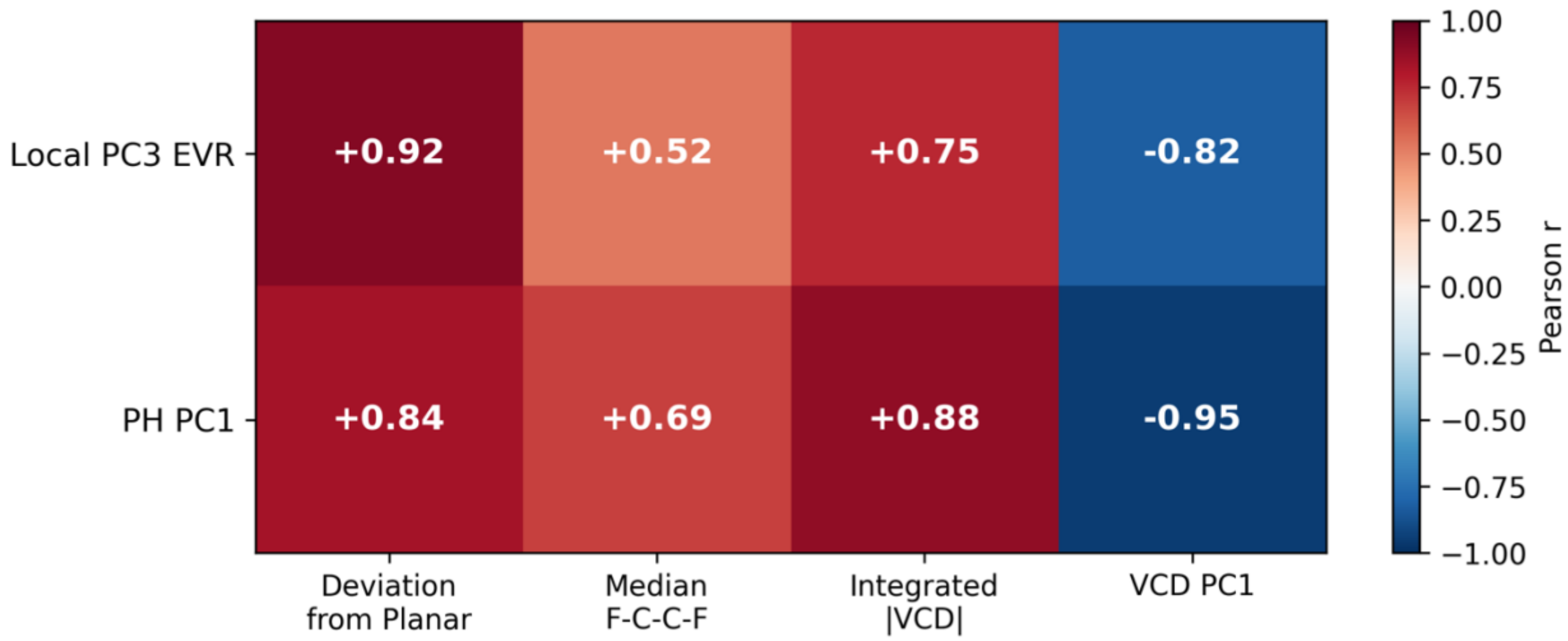

**Figure 4:** Pearson $r$ correlation matrix between two statistical helicity descriptors and four independent benchmarks, computed for the PFCA homologous series. Rows: Local PC3 EVR (local backbone PCA third component explained variance ratio); PH PC1 (persistent-homology first principal component). Columns: C-C-C-C deviation from planarity (geometric); median F-C-C-F dihedral (geometric, fluorinated molecules only); integrated absolute VCD intensity in the 1000 to 1500 $cm^{-1}$ C-F stretching region (spectroscopic); VCD PC1 from PCA on simulated VCD spectra restricted to 100 to 1500 $cm^{-1}$ (spectroscopic). Cell values are signed Pearson $r$; colormap runs from -1 (blue) to +1 (red). PH PC1 shows the strongest correlations across the median F-C-C-F dihedral and both spectroscopic benchmarks (integrated |VCD| and VCD PC1), and is comparable to Local PC3 EVR on the deviation-from-planarity benchmark, which saturates across the series. Spatial PC2 is excluded because it captures the class-level morphological dichotomy (Figure 2a) rather than within-class helicity gradients.

Although the statistical descriptors and both benchmark families are derived from the same DFT-optimized coordinates, the helical backbone they quantify is independently supported by experiment. The synchrotron powder crystal structure of perfluorononanoic acid ($^{F}C_9$) confirms that long-chain perfluoroalkanoic acids adopt a non-planar, twisted backbone in the solid state, and that study benchmarked its Rietveld-refined geometry against a DFT-optimized structure[35]; our independently optimized $^{F}C_9$ geometry reproduces the same helical backbone. Solid-state 19F MAS NMR of perfluoroalkanoic-acid assemblies[34] and studies of interfacial ordering of perfluorinated acids[36] are likewise consistent with an ordered helical perfluoroalkyl conformation. These experimental observations provide an external, non-DFT reference indicating that the helicity captured by the descriptors reflects a real structural feature rather than a peculiarity of the optimization protocol.

### 3.3. Spectroscopic validation with Vibrational Circular Dichroism (VCD)

Validation against the geometric benchmarks establishes internal consistency of the statistical framework. An independent test requires a direct physical observable sensitive to backbone chirality. VCD measures the differential absorption of left- and right-circularly polarized infrared radiation, providing exquisite sensitivity to rotationally chiral global three-dimensional molecular structure. Helicity in long-chain PFAS represents a form of axial chirality, making helical PFAS molecules VCD-active in proportion to their degree of helical twist[16]. DFT-calculated VCD spectra therefore provide a spectroscopic benchmark that is independent of the geometric measures used above.

We emphasize how this benchmark should be read physically. Conformational helicity here is P/M-degenerate and rapidly interconverting, so the VCD of a real racemic ensemble is zero by symmetry because the P and M helices contribute equal and opposite signals. The DFT VCD computed for a single helical sense is therefore not a prediction of an observable racemate spectrum; it is used as a computational chiroptical probe. Specifically, the magnitude of the single-sense response: integrated |VCD| and the amplitude captured by VCD PC1: grows monotonically with the helical amplitude of the conformer and thus serves as an independent, geometry-derived scalar for helicity that is orthogonal to the real-space dihedral benchmarks. It is this magnitude, not a signed observable, that we correlate against the descriptors; an experimentally observable single-sense signal would require a resolved or bias-induced helical population, e.g., in a chiral environment or at a surface.

For visual clarity, Figure 5 shows only the even-carbon chain members of the PFCA series; the complete odd-carbon series is provided in Supplementary Section S1 and follows the same progression. The spectral progression is physically clear: ${}^{F}C_2$ and ${}^{F}C_4$ show near-zero VCD intensity in the C-F stretching band (1000 to 1500 cm-1), consistent with their nascent helical character. ${}^{F}C_6$ displays the first weakly nonzero signal, and ${}^{F}C_8$ marks the onset of clearly nonzero chiroptical activity, which increases systematically through ${}^{F}C_{16}$. Hydrogenated analogues show negligible VCD signal throughout, confirming that the chiroptical activity is specific to the fluorinated backbone conformation. The onset at ${}^{F}C_8$ is consistent with the backbone dihedral metrics: below $C_8$, deviation from planarity remains below approximately 10° (Figure 3a), insufficient to generate a sustained chiral excess.

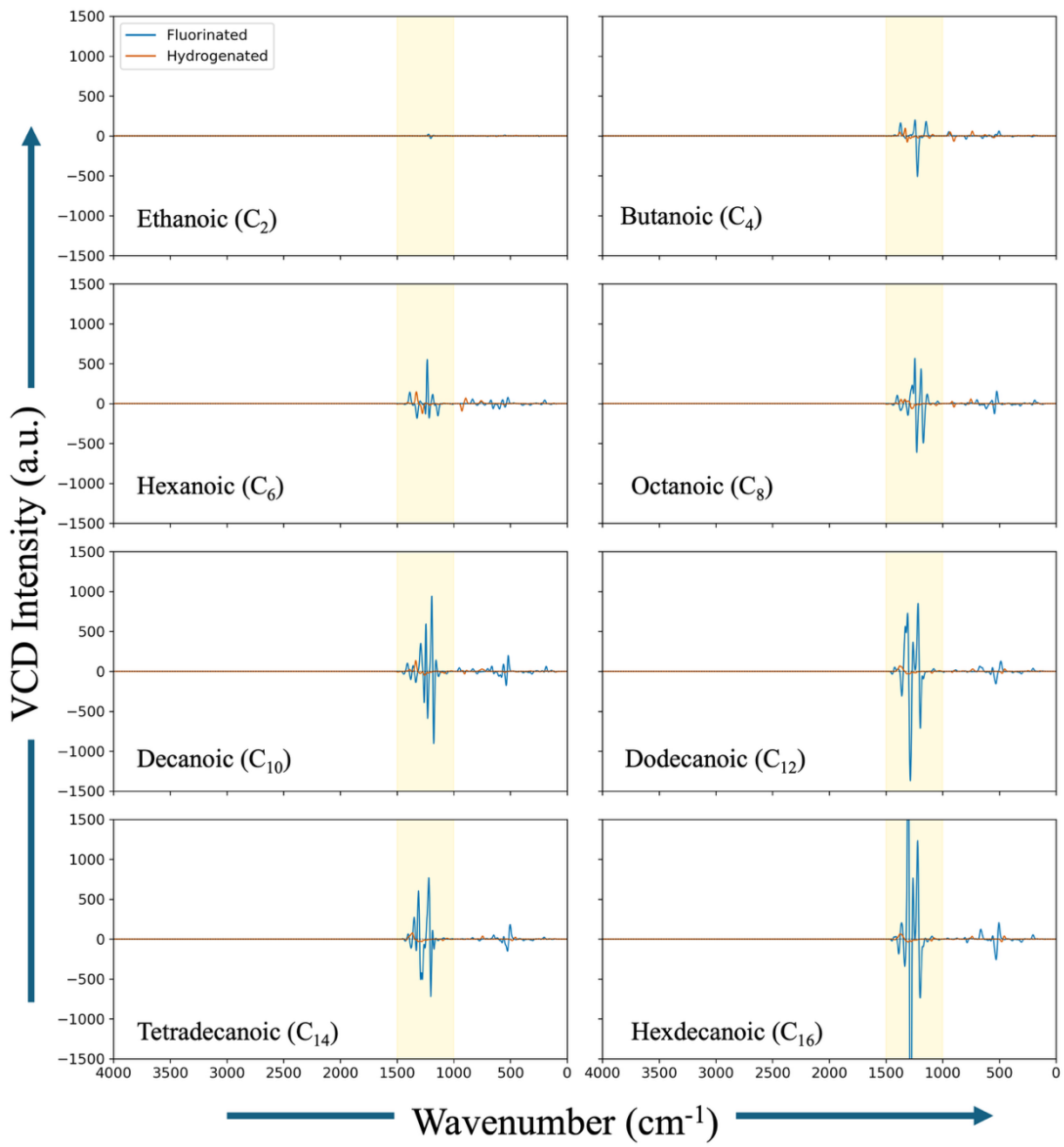

**Figure 5:** DFT-calculated VCD spectra for the even-carbon members of the PFCA homologous series; odd-carbon members are provided in Supplementary Section S1 for completeness. Panels progress from $C_2$ (top left) through $C_{16}$ (bottom right). Blue lines: perfluorinated carboxylic acids. Red lines: hydrogenated analogues. Wavenumber axes follow standard spectroscopic convention, running from 4000 $cm^{-1}$ at the left to 0 $cm^{-1}$ at the right; the shaded region marks the C-F stretching band at 1000-1500 $cm^{-1}$. $^FC_2$ and $^FC_4$ show near-zero VCD response; $^FC_6$ shows weak activity; $^FC_8$ marks the onset of clearly nonzero chiroptical signal, which increases systematically through $^FC_{16}$. Hydrogenated analogues show negligible signal throughout. Panels share x and y axis scales. VCD is computed for a single helical sense as a probe of chiroptical magnitude; the racemic ensemble is P/M-degenerate and its net VCD is zero by symmetry, so the panels report the magnitude of the single-sense response rather than an observable racemate spectrum.

To make the spectroscopic validation visually explicit, Figure 6 shows scatter plots of PH PC1 and Local PC3 EVR against VCD PC1 for the PFCA series. Both descriptors correlate strongly with VCD PC1 (the sign is a PCA-axis convention), with PH PC1 showing a notably tighter relationship (Pearson $r$ = -0.95) than Local PC3 EVR ($r$ = -0.82). This visual confirmation, alongside Figure 4, demonstrates that the statistical descriptors capture the same physical phenomenon: progressive helical twist, that manifests in the chiroptical response. Spatial PC2 is excluded from Figure 4 because it resolves the class-level fluorinated/hydrogenated dichotomy (Figure 2a) but does not capture the within-class chain-length helicity gradient; its role is therefore distinct from Local PC3 EVR and PH PC1, which both track progressive helical twist within the fluorinated series.

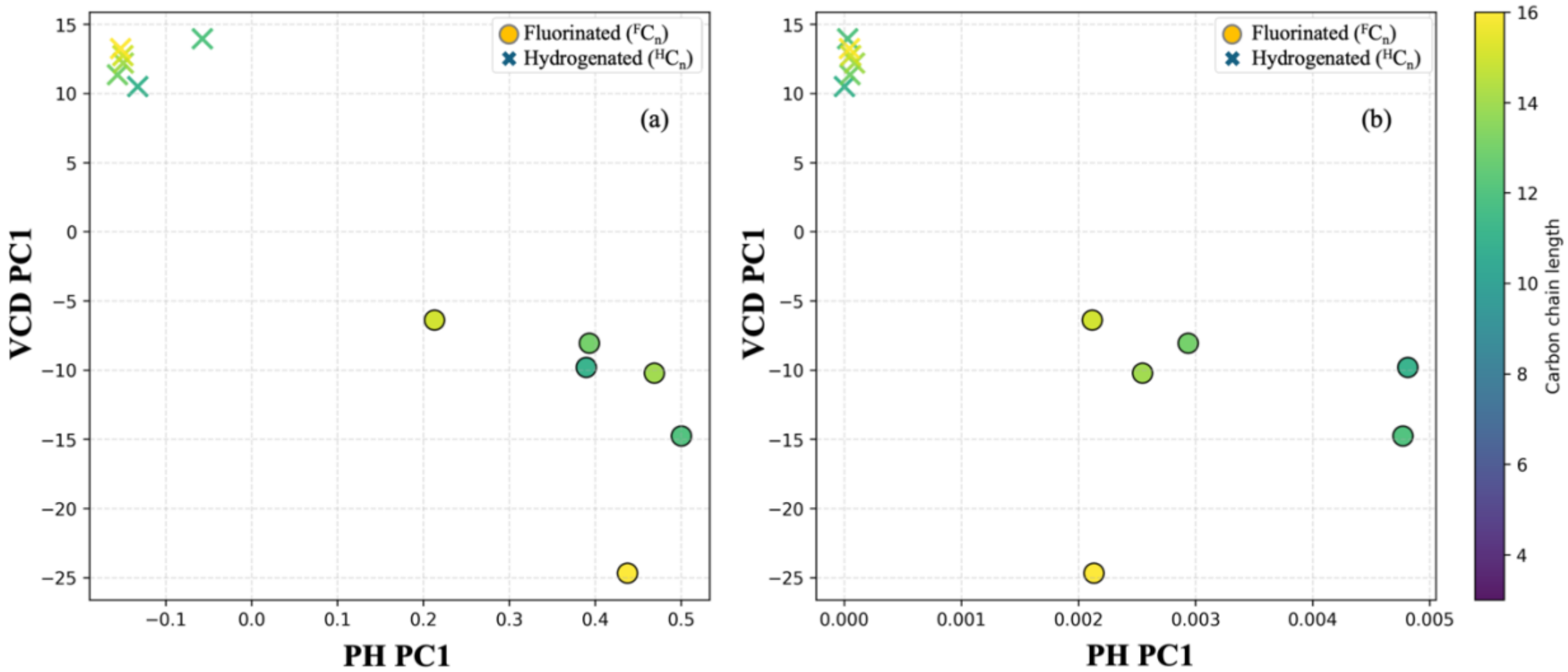

**Figure 6:** Scatter plots of statistical descriptors versus VCD PC1 for the PFCA series. (a) PH PC1 versus VCD PC1 (Pearson r = -0.95). (b) Local PC3 EVR versus VCD PC1 (Pearson r = -0.82). Points are colored by carbon chain length. VCD PC1 is the first principal component of PCA on the simulated VCD spectra restricted to 100 to 1500 $cm^{-1}$, capturing the dominant spectral variation across the series. Both descriptors correlate strongly with VCD PC1 (the negative sign is a PCA convention for the VCD PC1 axis; magnitude is the point), with PH PC1 showing the tighter relationship.

Two scalar summaries of the VCD response are extracted for quantitative comparison with the statistical descriptors. The integrated absolute VCD intensity in the 1000-1500 $cm^{-1}$ C-F stretching band directly quantifies the magnitude of the chiroptical signal per molecule. The first principal component of PCA applied to the full VCD spectra restricted to 100-1500 $cm^{-1}$ (VCD PC1) captures the dominant statistical mode of spectral variation across the series. Both are reported in the Pearson $r$ correlation matrix (Figure 4). PH PC1 shows the strongest Pearson correlations with both VCD summaries within the PFCA series, followed by Local PC3 EVR and then Spatial PC2. This ordering is consistent with the geometric benchmark pattern: the descriptor that most directly targets the out-of-plane topology of the helix (PH PC1, via arc-length normalization) is the strongest helicity proxy across both geometric and spectroscopic benchmarks. The convergence of three independent statistical descriptors with two geometric and two spectroscopic benchmarks establishes that the quantified gradients in backbone geometry are real, physically significant structural phenomena.

### 3.4. Generalizability across fixed-chain-length subfamilies with varying fluorine content

The PFCA series provides a clean testbed because helical character scales monotonically with chain length, making the ordering of descriptor values straightforward to interpret. A more demanding test of robustness is to apply the same descriptors to subfamilies where chain length is fixed and structural variation arises instead from differences in fluorine substitution pattern and headgroup chemistry. The PFOA analogue set (Figure 7) comprising partially to fully fluorinated octanoyl-chain molecules of fixed carbon backbone length, and the PFOS analogue set (Figure 8), the equivalent fixed-length series with a sulfonic acid headgroup, provide exactly this test: within each set, all molecules share the same carbon backbone length, and variation in the number and arrangement of fluorine atoms drives differences in backbone conformation.

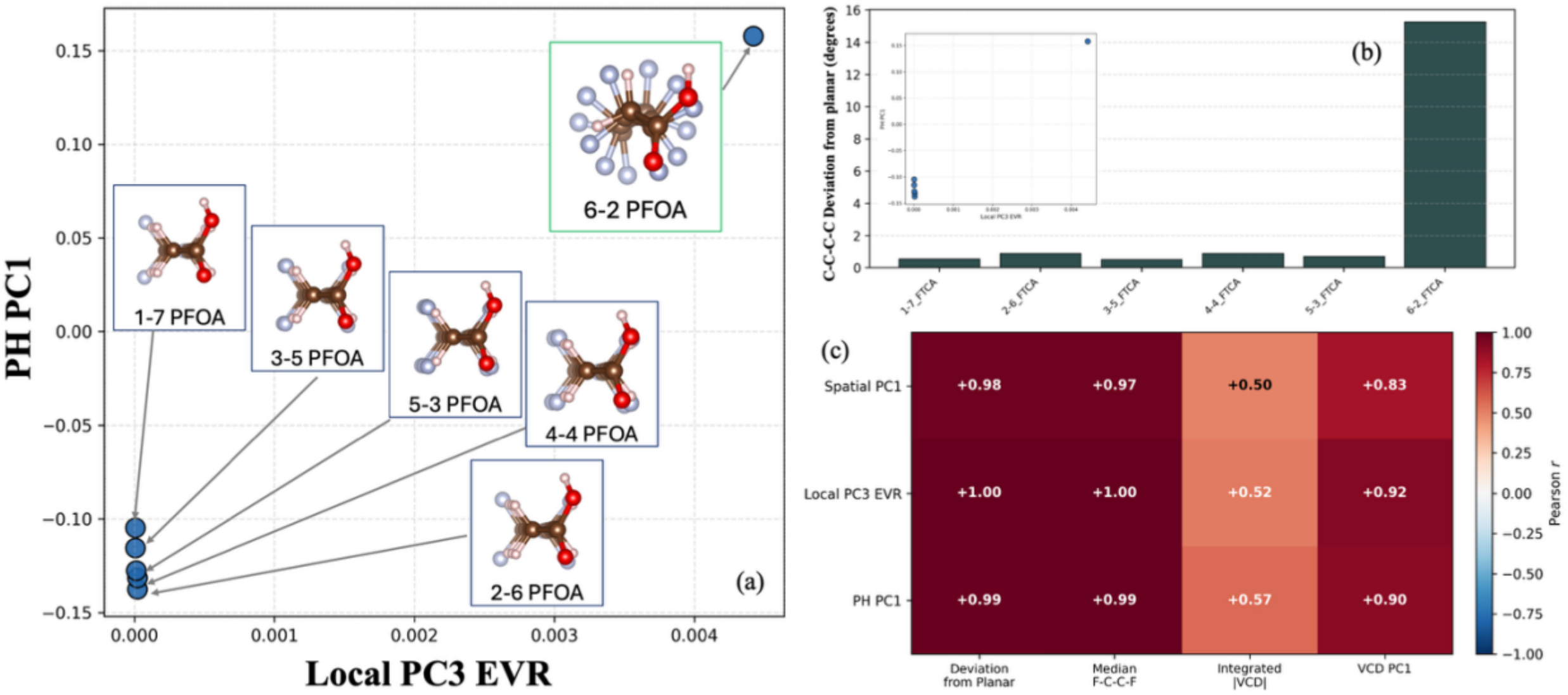

**Figure 7:** Statistical descriptors and geometric benchmarks for the PFOA subfamily (fixed chain length). (a) Local PC3 EVR versus PH PC1. (b) C-C-C-C backbone dihedral deviation from planarity for each PFOA molecule; bars are shown in a single color because chain length does not vary within this subfamily. A non-annotated version of plot (a) is showed as the inlay, (c) Pearson $r$ correlation matrix with three descriptor rows (Spatial PC1, Local PC3 EVR, PH PC1) and four benchmark columns (C-C-C-C deviation from planarity; median F-C-C-F dihedral; integrated absolute VCD intensity in the 1000 to 1500 $cm^{-1}$ C-F stretching region; VCD PC1). Spatial PC1 is included here because within a fixed-chain-length subfamily it encodes structural variation beyond simple chain extent. All panels are shown without point annotations.

The inset structures in Figures 7a and 8a illustrate the molecular diversity within each set, from partially fluorinated to fully perfluorinated architectures. To provide validation independent of visual structural inspection, the Pearson $r$ correlation matrices in Figures 7c and 8c include spectroscopic benchmarks: integrated absolute VCD intensity in the 1000-1500 $cm^{-1}$ C-F stretching region and VCD PC1. The strong correlations of Local PC3 EVR and PH PC1 with these VCD benchmarks confirm that the statistical descriptors capture genuine conformational differences driven by fluorine substitution pattern rather than chain-length artifacts; full per-molecule VCD spectra for both subfamilies are provided in Supplementary Section S1. In Figure 7a, the scatter of Local PC3 EVR versus PH PC1 shows that more heavily fluorinated PFOA analogues occupy higher values of both descriptors, consistent with greater helical character. Figure 7b confirms that the C-C-C-C backbone dihedral deviation from planarity tracks this trend: molecules with higher fluorine content show larger deviations. The Pearson $r$ correlation matrix (Figure 7c) demonstrates that Spatial PC1, Local PC3 EVR, and PH PC1 all correlate with the geometric and spectroscopic benchmarks within this fixed-chain-length set, with PH PC1 showing the strongest correlations across the median-dihedral and both spectroscopic benchmarks and being comparable to Local PC3 EVR on deviation-from-planarity. Critically, Spatial PC1, not PC2, is the informative spatial autocorrelation descriptor here, because within a fixed-chain-length subfamily PC1 encodes the cross-sectional morphological differences driven by fluorine substitution rather than being dominated by chain extent.

The PFOS analogue set (Figure 8) extends this observation to a structurally distinct headgroup chemistry: the sulfonic acid group and confirms the same pattern. The correlation structure in Figure 8c mirrors that of Figure 7c, and the inset structures in Figure 8a again show that higher fluorine content is associated with larger Local PC3 EVR and PH PC1 values. Together, Figures 7 and 8 establish that the statistical descriptors detect the helicity signature of fluorine substitution independently of chain length, and that the descriptor hierarchy established for PFCAs generalizes to chemically diverse PFAS subclasses including novel perfluorosulfonic acid architectures. This is the key result enabling these descriptors to be used in screening exercises where PFAS structures vary in both chain length and fluorine substitution simultaneously.

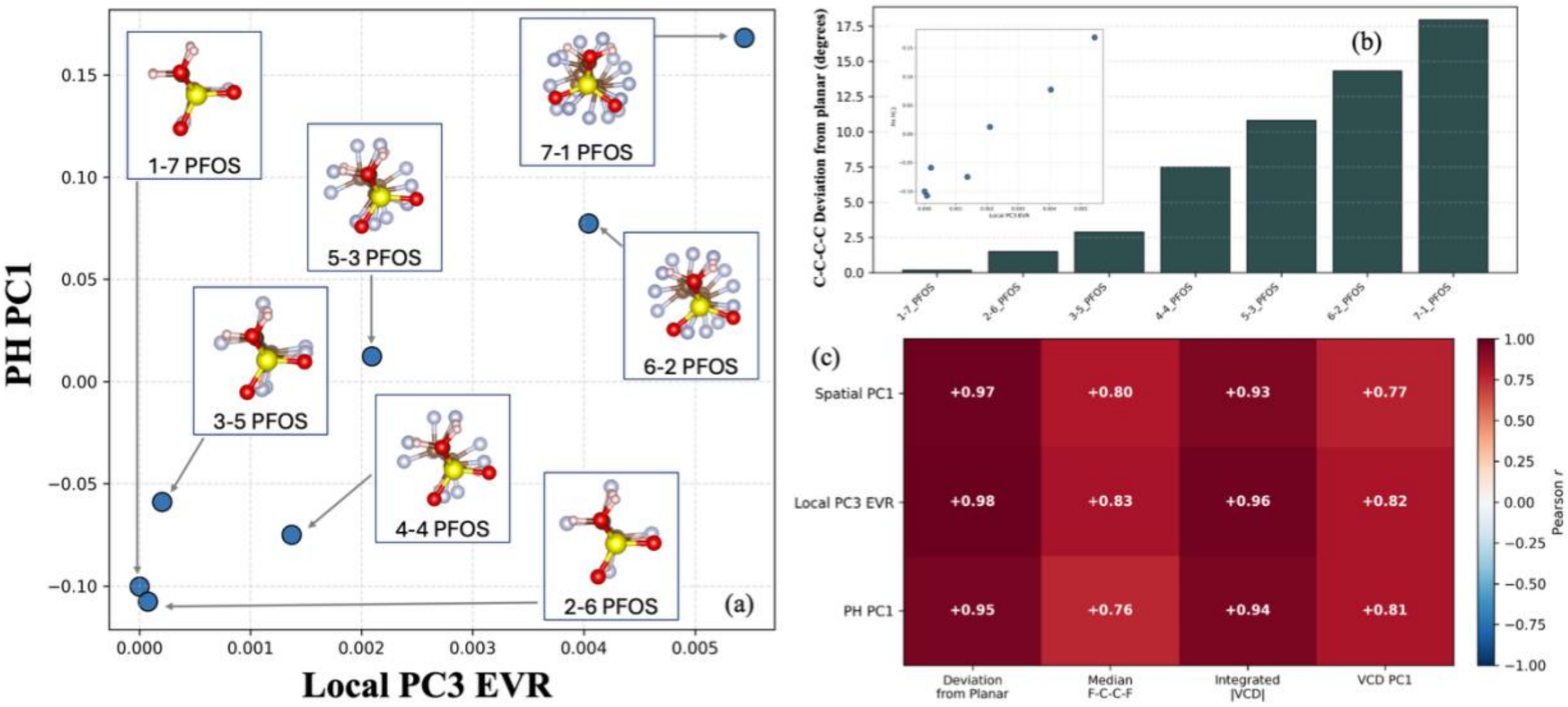

**Figure 8:** Statistical descriptors and geometric benchmarks for the PFOS subfamily. Panels follow the same layout as Figure 7: (a) Local PC3 EVR versus PH PC1, (b) C-C-C-C backbone dihedral deviation from planarity, a non-annotated version of plot (a) is showed as the inlay, and (c) Pearson $r$ correlation matrix with Spatial PC1, Local PC3 EVR, and PH PC1 as rows and the four benchmarks as columns. All panels are shown without point annotations.

## 4. CONCLUSIONS

PH PC1 from persistent homology of the arc-length-normalized backbone is the most physically faithful continuous helicity descriptor for PFAS, showing the strongest Pearson correlations with the median F-C-C-F dihedral and both spectroscopic benchmarks (integrated VCD intensity in the 1000 to 1500 $cm^{-1}$ C-F stretching region and VCD PC1), and correlations comparable to Local PC3 EVR against backbone dihedral deviation from planarity, in both the PFCA homologous series and the expanded multi-class dataset. Spatial PC2 separates fluorinated from hydrogenated classes by backbone morphology. Local PC3 EVR provides a per-molecule measure of out-of-plane backbone variance that is systematically larger for fluorinated chains than for their hydrogenated analogues at each chain length; as an explained-variance ratio it is largest for short-to-mid chains and does not increase monotonically with chain length, functioning as a class discriminator rather than a within-series gradient.

Our framework reveals that helical character in PFAS is not a binary property but exists on a continuum. Nascent backbone twist is detectable from $^{F}C_3$ onward in both the C-C-C-C dihedral and F-C-C-F benchmarks, consistent with the established gauche character of short perfluoroalkyl chains[13-15]. The statistical descriptors resolve a progressive increase in helical character with chain elongation, while the chiroptical VCD signal becomes clearly nonzero from $^{F}C_8$ onward, where the backbone dihedral deviation from planarity exceeds approximately 10° and a sustained chiral excess is established. The quantitative agreement of the geometric and topological measures across PFCA, PFSA, PFHS, FTOH, and PFOA/PFOS subfamilies confirms that the helical twist is a general property of the perfluoroalkyl chain rather than a peculiarity of any single headgroup chemistry. The fixed-chain-length PFOA and PFOS analogue sets further demonstrate that the statistical descriptors resolve helicity differences driven by fluorine substitution pattern alone, without variation in chain length: a property essential for screening structurally diverse PFAS libraries where both chain length and fluorination degree vary simultaneously.

A validated, continuous helicity descriptor can now be incorporated into QSPR models linking three-dimensional backbone conformation to environmental fate, bioavailability, and degradation reactivity. PH PC1, particularly, provides a structure-based handle for computationally pre-screening catalytic substrates and reaction conditions: including those that facilitate electron transfer: that promote the helical-to-linear conformational transition proposed to precede defluorination, offering a rational pathway toward more effective remediation strategies. We stress that the coupling between this conformational transition and defluorination is a computational hypothesis; the value of the descriptor for screening does not depend on the precise molecular mechanism by which linearization facilitates C-F scission.

More broadly, this study demonstrates that treating molecules as three-dimensional data objects and applying feature engineering techniques from materials informatics and topological data analysis reveals fundamental structure-property relationships without reliance on traditional, human-defined chemical descriptors. This approach is well-suited to molecular systems that exhibit intrinsic helicity arising from their bonding chemistry, such as fluorinated chains and synthetic helical polymers, and can be extended to other classes of persistent organic pollutants and functional materials where backbone conformation governs environmental fate or reactivity. All DFT-optimized structures, backbone coordinates, and descriptor calculation codes used in this study are openly available at the repository listed in the Data Availability Statement, facilitating direct extension of this framework to other PFAS subfamilies and persistent organic pollutant classes. We note that the present descriptors are evaluated on single minimum-energy conformers; Boltzmann-weighted conformer ensembles, which the descriptors natively support, are a clear direction for extending the framework to flexible or partially fluorinated chains where multiple conformers are thermally accessible.

## 5. DATA AND SOFTWARE AVAILABILITY

The data and codes used for this study is publicly available at: https://doi.org/10.5281/zenodo.22018077. Alternatively, the data is also available at: https://github.com/pranoy-ray/PFAShelix

## 6. ACKNOWLEDGEMENTS

PR and SK acknowledge support from NSF DMREF Award 2119640. MKV acknowledges OUSD for the FY24-26 LUCI Fellowship. SV and HC acknowledge financial support from the National Science Foundation (CHE-2109210) and Army Corps of Engineers (cooperative agreement - W912HZ-23-2-0009). SV and HC also gratefully acknowledge the allocated computational resources at the High-Performance Computing facility at Colorado School of Mines.

Any opinion, findings, and conclusions or recommendations expressed in this material are those of the author(s) and do not necessarily reflect the views of the National Science Foundation.

## 7. SUPPLEMENTARY INFORMATION

The supplementary information is available free of charge at the online version of this article and contains DFT computational details; Additional VCD spectra; Persistent homology detailed procedures; Additional persistent homology figures and parameterizations; C-C-C-C dihedral backbone measurements; Poly-fluorinated PFOA, PFOS, PFSA, and PFHS VCD spectra and corresponding dihedral analysis; Pearson *r* correlation matrices for PCA; Additional PC vs. dihedral figures.

## 8. REFERENCES

(1) Kolel-Veetil, M. The Promise of PFAS Remediation by the Fourth State of Matter. *Curr. Opin. Chem. Eng.* **2024**, *43*, 100982. https://doi.org/10.1016/j.coche.2023.100982.
(2) Capodaglio, A. G. Prospects of Novel Technologies for PFAS Destruction in Water and Wastewater. *Appl. Sci.* **2025**, *15* (17). https://doi.org/10.3390/app15179311.
(3) Rahman, M. H.-U.; Sikder, R.; Tonmoy, T. A.; Hossain, Md. M.; Ye, T.; Aich, N.; Gadhamshetty, V. Transforming PFAS Management: A Critical Review of Machine Learning Applications for Enhanced Monitoring and Treatment. *J. Water Process Eng.* **2025**, *70*, 106941. https://doi.org/10.1016/j.jwpe.2025.106941.
(4) Verma, S.; Lee, T.; Sahle-Demessie, E.; Ateia, M.; Nadagouda, M. N. Recent Advances on PFAS Degradation *via* Thermal and Nonthermal Methods. *Chem. Eng. J. Adv.* **2023**, *13*, 100421. https://doi.org/10.1016/j.ceja.2022.100421.

(5) Rao, U.; Su, Y.; Khor, C. M.; Jung, B.; Ma, S.; Cwiertny, D. M.; Wong, B. M.; Jassby, D. Structural Dependence of Reductive Defluorination of Linear PFAS Compounds in a UV/Electrochemical System. *Environ. Sci. Technol.* **2020**, *54* (17), 10668–10677. https://doi.org/10.1021/acs.est.0c02773.
(6) *CompTox Chemicals Dashboard*. https://comptox.epa.gov/dashboard/chemical-lists/pfasstruct (accessed 2025-09-09).
(7) Das, S.; Ronen, A.; Das, S.; Ronen, A. A Review on Removal and Destruction of Per- and Polyfluoroalkyl Substances (PFAS) by Novel Membranes. *Membranes* **2022**, *12* (7). https://doi.org/10.3390/membranes12070662.
(8) Hamza, M.; Ayinla, R. T.; Elsayed, I.; Hassan, E. B.; Hamza, M.; Ayinla, R. T.; Elsayed, I.; Hassan, E. B. Understanding PFAS Adsorption: How Molecular Structure Affects Sustainable Water Treatment. *Environments* **2025**, *12* (9). https://doi.org/10.3390/environments12090330.
(9) Fenton, S. E.; Ducatman, A.; Boobis, A.; DeWitt, J. C.; Lau, C.; Ng, C.; Smith, J. S.; Roberts, S. M. Per- and Polyfluoroalkyl Substance Toxicity and Human Health Review: Current State of Knowledge and Strategies for Informing Future Research. *Environ. Toxicol. Chem.* **2021**, *40* (3), 606–630. https://doi.org/10.1002/etc.4890.
(10) Mudlaff, M.; Sosnowska, A.; Gorb, L.; Bulawska, N.; Jagiello, K.; Puzyn, T. Environmental Impact of PFAS: Filling Data Gaps Using Theoretical Quantum Chemistry and QSPR Modeling. *Environ. Int.* **2024**, *185*, 108568. https://doi.org/10.1016/j.envint.2024.108568.
(11) Yin, S.; Calvillo Solís, J. J.; Sandoval-Pauker, C.; Puerto-Diaz, D.; Villagrán, D. Advances in PFAS Electrochemical Reduction: Mechanisms, Materials, and Future Perspectives. *J. Hazard. Mater.* **2025**, *491*, 137943. https://doi.org/10.1016/j.jhazmat.2025.137943.
(12) Li, G.; Peng, M.; Huang, Q.; Huang, C.-H.; Chen, Y.; Hawkins, G.; Li, K. A Review on the Recent Mechanisms Investigation of PFAS Electrochemical Oxidation Degradation: Mechanisms, DFT Calculation, and Pathways. *Front. Environ. Eng.* **2025**, *4*. https://doi.org/10.3389/fenve.2025.1568542.
(13) Verma, S.; Lee, T.; Sahle-Demessie, E.; Ateia, M.; Nadagouda, M. N. Recent Advances on PFAS Degradation *via* Thermal and Nonthermal Methods. *Chem. Eng. J. Adv.* **2023**, *13*, 100421. https://doi.org/10.1016/j.ceja.2022.100421.
(14) Biswas, S.; Wong, B. M. Beyond Conventional Density Functional Theory: Advanced Quantum Dynamical Methods for Understanding Degradation of Per- and Polyfluoroalkyl Substances. *ACS EST Eng.* **2023**, *4* (1), 96–104. https://doi.org/10.1021/acsestengg.3c00216.
(15) Bunn, C. W.; Howells, E. R. Structures of Molecules and Crystals of Fluoro-Carbons. *Nature* **1954**, *174* (4429), 549–551. https://doi.org/10.1038/174549a0.
(16) Cormanich, R. A.; O'Hagan, D.; Bühl, M. Hyperconjugation Is the Source of Helicity in Perfluorinated N-Alkanes. *Angew. Chem. Int. Ed.* **2017**, *56* (27), 7867–7870. https://doi.org/10.1002/anie.201704112.
(17) McTaggart, M.; Malardier-Jugroot, C. The Role of Helicity in PFAS Resistance to Degradation: DFT Simulation of Electron Capture and Defluorination. *Phys. Chem. Chem. Phys.* **2024**, *26* (5), 4692–4701. https://doi.org/10.1039/D3CP04973F.
(18) Jang, S. S.; Blanco, M.; Goddard, W. A.; Caldwell, G.; Ross, R. B. The Source of Helicity in Perfluorinated N-Alkanes. *Macromolecules* **2003**, *36* (14), 5331–5341. https://doi.org/10.1021/ma025645t.

(19) Mifkovic, M.; Van Hoomissen, D. J.; Vyas, S. Conformational Distributions of Helical Perfluoroalkyl Substances and Impacts on Stability. *J. Comput. Chem.* **2022**, *43* (24), 1656–1661. https://doi.org/10.1002/jcc.26967.
(20) Kujawa, J.; Cerneaux, S.; Kujawski, W. Characterization of the Surface Modification Process of Al2O3, TiO2 and ZrO2 Powders by PFAS Molecules. *Colloids Surf. Physicochem. Eng. Asp.* **2014**, *447*, 14–22. https://doi.org/10.1016/j.colsurfa.2014.01.065.
(21) *Potential-Driven Electron Transfer Lowers the Dissociation Energy of the C–F Bond and Facilitates Reductive Defluorination of Perfluorooctane Sulfonate (PFOS) | ACS Applied Materials & Interfaces*. https://pubs.acs.org/doi/10.1021/acsami.9b10449 (accessed 2025-09-09).
(22) *What can Blyholder teach us about PFAS degradation on metal surfaces? - Environmental Science: Advances (RSC Publishing)*. https://pubs.rsc.org/en/content/articlelanding/2024/va/d3va00281k (accessed 2025-09-09).
(23) Lu, L.; Na, C. Halogen Bonding in Perfluoroalkyl Adsorption. *ACS Omega* **2024**, *9* (24), 26050–26057. https://doi.org/10.1021/acsomega.4c01367.
(24) *Structural Dependence of Reductive Defluorination of Linear PFAS Compounds in a UV/Electrochemical System | Environmental Science & Technology*. https://pubs.acs.org/doi/10.1021/acs.est.0c02773 (accessed 2025-09-09).
(25) Nayak, S. K.; Bhardwaj, K.; Verma, P. K.; Yamijala, S. S. R. K. C. Plasmon-Induced Degradation of Short-Chain PFAS by Noble Metal Nanoclusters. *J. Phys. Chem. Lett.* **2025**, *16* (31), 8046–8055. https://doi.org/10.1021/acs.jpclett.5c01683.
(26) Brusseau, M. L. The Influence of Molecular Structure on the Adsorption of PFAS to Fluid-Fluid Interfaces: Using QSPR to Predict Interfacial Adsorption Coefficients. *Water Res.* **2019**, *152*, 148–158. https://doi.org/10.1016/j.watres.2018.12.057.
(27) Roy, K.; Kar, S.; Das, R. N. *A Primer on QSAR/QSPR Modeling: Fundamental Concepts*; Springer, 2015.
(28) Ray, P.; Choudhary, K.; Kalidindi, S. R. Lean CNNs for Mapping Electron Charge Density Fields to Material Properties. *Integrating Mater. Manuf. Innov.* **2025**, *14* (1), 1–13. https://doi.org/10.1007/s40192-024-00389-9.
(29) Kaundinya, P. R.; Choudhary, K.; Kalidindi, S. R. Machine Learning Approaches for Feature Engineering of the Crystal Structure: Application to the Prediction of the Formation Energy of Cubic Compounds. *Phys. Rev. Mater.* **2021**, *5* (6), 063802. https://doi.org/10.1103/PhysRevMaterials.5.063802.
(30) Ray, P.; Generale, A. P.; Vankireddy, N.; Asoma, Y.; Nakauchi, M.; Lee, H.; Yoshida, K.; Okuno, Y.; Kalidindi, S. R. Refining Coarse-Grained Molecular Topologies: A Bayesian Optimization Approach. *Npj Comput. Mater.* **2025**, *11* (1), 234. https://doi.org/10.1038/s41524-025-01729-9.
(31) Ray, P.; Yuichiro, A.; Vankireddy, N.; Generale, A. P.; Masataka, N.; Haein, L.; Katsuhisa, Y.; Kalidindi, S. R.; Yoshishige, O. Assessing the Accuracy of Bayesian-Optimized CGMD in Predicting Polymer Miscibility. ChemRxiv December 1, 2025. https://doi.org/10.26434/chemrxiv-2025-p14dx.
(32) Ray, P.; Castillo, A. R.; Kolel-Veetil, M.; Kalidindi, S. R. ML Workflows for Screening Degradation-Relevant Properties of Forever Chemicals. *Adv. Sci.* **2026**, *n/a* (n/a), e23817. https://doi.org/10.1002/advs.202523817.

(33) Monde, K.; Miura, N.; Hashimoto, M.; Taniguchi, T.; Inabe, T. Conformational Analysis of Chiral Helical Perfluoroalkyl Chains by VCD. *J. Am. Chem. Soc.* **2006**, *128* (18), 6000–6001. https://doi.org/10.1021/ja0602041.

(34) Pawsey, S.; Reven, L. 19F Fast Magic-Angle Spinning NMR Studies of Perfluoroalkanoic Acid Self-Assembled Monolayers. *Langmuir* **2006**, *22* (3), 1055–1062. https://doi.org/10.1021/la051725p.

(35) Reid, J. W.; Malik, T.; Pravica, M. G.; Leontowich, A. F. G.; Rahemtulla, A. Crystal Structure of Perfluorononanoic Acid, C9HF17O2. *Powder Diffr.* **2024**, *39* (4), 263–269. https://doi.org/10.1017/S0885715624000356.

(36) Schwidetzky, R.; Sun, Y.; Fröhlich-Nowoisky, J.; Kunert, A. T.; Bonn, M.; Meister, K. Ice Nucleation Activity of Perfluorinated Organic Acids. *J. Phys. Chem. Lett.* **2021**, *12* (13), 3431–3435. https://doi.org/10.1021/acs.jpclett.1c00604.

(37) Hohenberg, P.; Kohn, W. Inhomogeneous Electron Gas. *Phys. Rev.* **1964**, *136* (3B), B864–B871. https://doi.org/10.1103/PhysRev.136.B864.

(38) Kohn, W.; Sham, L. J. Self-Consistent Equations Including Exchange and Correlation Effects. *Phys. Rev.* **1965**, *140* (4A), A1133–A1138. https://doi.org/10.1103/PhysRev.140.A1133.

(39) Chakraborty, B.; Ray, P.; Garg, N.; Banerjee, S. High Capacity Reversible Hydrogen Storage in Titanium Doped 2D Carbon Allotrope Ψ-Graphene: Density Functional Theory Investigations. *Int. J. Hydrog. Energy* **2021**, *46* (5), 4154–4167. https://doi.org/10.1016/j.ijhydene.2020.10.161.

(40) Nair, H. T.; Kundu, A.; Ray, P.; Jha, P. K.; Chakraborty, B. Ti-Decorated C30 as a High-Capacity Hydrogen Storage Material: Insights from Density Functional Theory. *Sustain. Energy Fuels* **2023**. https://doi.org/10.1039/D3SE00845B.

(41) Kundu, A.; Jaiswal, A.; Ray, P.; Sahu, S.; Chakraborty, B. Zr Doped C $_{24}$ Fullerene as Efficient Hydrogen Storage Material: Insights from DFT Simulations. *J. Phys. Appl. Phys.* **2024**, *57* (49), 495502. https://doi.org/10.1088/1361-6463/ad75a1.

(42) Neese, F. The ORCA Program System. *WIREs Comput. Mol. Sci.* **2012**, *2* (1), 73–78. https://doi.org/10.1002/wcms.81.

(43) Hehre, W. J.; Ditchfield, R.; Pople, J. A. Self—Consistent Molecular Orbital Methods. XII. Further Extensions of Gaussian—Type Basis Sets for Use in Molecular Orbital Studies of Organic Molecules. *J. Chem. Phys.* **1972**, *56* (5), 2257–2261. https://doi.org/10.1063/1.1677527.

(44) Hariharan, P. C.; Pople, J. A. The Influence of Polarization Functions on Molecular Orbital Hydrogenation Energies. *Theor. Chim. Acta* **1973**, *28* (3), 213–222. https://doi.org/10.1007/BF00533485.

(45) Giroday, T.; Montero-Campillo, M. M.; Mora-Diez, N. Thermodynamic Stability of PFOS: M06-2X and B3LYP Comparison. *Comput. Theor. Chem.* **2014**, *1046*, 81–92. https://doi.org/10.1016/j.comptc.2014.08.003.

(46) Hidalgo, A.; Giroday, T.; Mora-Diez, N. Thermodynamic Stability of Neutral and Anionic PFOAs. *Theor. Chem. Acc.* **2015**, *134* (11), 124. https://doi.org/10.1007/s00214-015-1725-4.

(47) Ray, P.; Bhowmik, S.; Suryanarayana, P.; Kalidindi, S. R.; Medford, A. J. Electronic Manifolds for Extrapolative Alloy Discovery. *Digit. Discov.* **2026**. https://doi.org/10.1039/D6DD00105J.

(48) Edelsbrunner; Letscher; Zomorodian. Topological Persistence and Simplification. *Discrete Comput. Geom.* **2002**, *28* (4), 511–533. https://doi.org/10.1007/s00454-002-2885-2.

(49) Bizana, G. B.; Kalidindi, S. R.; Barrales-Mora, L. A. Data-Driven Quantification of Grain Boundary Structures and Their Correlation with Migration Kinetics. Social Science Research Network: Rochester, NY March 27, 2026. https://doi.org/10.2139/ssrn.6477786.
(50) Obayashi, I.; Hiraoka, Y.; Kimura, M. Persistence Diagrams with Linear Machine Learning Models. *J. Appl. Comput. Topol.* **2018**, *1* (3), 421–449. https://doi.org/10.1007/s41468-018-0013-5.
(51) Obayashi, I.; Nakamura, T.; Hiraoka, Y. Persistent Homology Analysis for Materials Research and Persistent Homology Software: HomCloud. *J. Phys. Soc. Jpn.* **2022**, *91* (9), 091013. https://doi.org/10.7566/JPSJ.91.091013.
(52) Ichinomiya, T.; Obayashi, I.; Hiraoka, Y. Protein-Folding Analysis Using Features Obtained by Persistent Homology. *Biophys. J.* **2020**, *118* (12), 2926–2937. https://doi.org/10.1016/j.bpj.2020.04.032.
(53) Adams, H.; Emerson, T.; Kirby, M.; Neville, R.; Peterson, C.; Shipman, P.; Chepushtanova, S.; Hanson, E.; Motta, F.; Ziegelmeier, L. Persistence Images: A Stable Vector Representation of Persistent Homology. *J Mach Learn Res* **2017**, *18* (1), 218–252.
(54) Landrum, G.; Tosco, P.; Kelley, B.; Rodriguez, R.; Cosgrove, D.; Vianello, R.; sriniker; Gedeck, P.; Jones, G.; Kawashima, E.; NadineSchneider; Nealschneider, D.; tadhurst-cdd; Dalke, A.; Swain, M.; Cole, B.; Turk, S.; Savelev, A.; Maeder, N.; Walker, R.; Vaucher, A.; Wójcikowski, M.; Faara, H.; Take, I.; Scalfani, V. F.; Pechersky, Y.; Ujihara, K.; Probst, D.; Monat, J.; Lehtivarjo, J. Rdkit/Rdkit: 2026_03_1 (Q1 2026) Release, 2026. https://doi.org/10.5281/zenodo.19250388.